\documentclass[acus]{JAC2003}
\usepackage[dvipdfm]{graphicx}
\usepackage{booktabs}
\usepackage{bm}
\usepackage{amsmath}
\usepackage{color}

\begin{document}
\title{
FAST VACUUM DECAY INTO PARTICLE PAIRS \\
IN STRONG ELECTRIC AND MAGNETIC FIELDS
\thanks{This work was supported by the Grant-in-Aid for the Global COE Program ``The Next Generation of Physics, Spun from Universality and Emergence'' from the Ministry of Education, Culture, Sports, Science and Technology (MEXT) of Japan.}
}

\author{Y. Hidaka, T. Iritani, and H. Suganuma, \\
Department of Physics, 
Kyoto University, Kitashirakawa Oiwakecho, Sakyo, Kyoto 606-8502, Japan
}

\maketitle

\begin{abstract}
We discuss fermion pair productions in strong electric and magnetic fields. 
We point out that, in the case of massless fermions, 
the vacuum persistency probability per unit time and volume is 
zero in the strong electric and magnetic fields, 
while it is finite when the magnetic field is absent.
The contribution from the lowest Landau level (LLL) dominates this phenomenon.
We also discuss dynamics of the vacuum decay, 
using an effective theory of the LLL projection, 
taking into account the back reaction.
\end{abstract}

\section{Introduction}
Dynamics in strong fields has been an interesting subject 
in theoretical physics. 
Recently, this subject is being paid attention also in
the experimental physics of creation of the quark gluon plasma.
In high-energy heavy-ion collision experiments, 
at the so-called Glasma stage~\cite{Glasma} just after the collision,
longitudinal color electric and magnetic fields are expected to be 
produced in the context of the color glass condensate 
of order $1$--$2$ GeV in RHIC and $5$ GeV in LHC.
In the peripheral collision, a strong magnetic field of order $100$ MeV 
would be induced. The question is how the strong fields decay 
and the system is thermalized.

In this work, we concentrate on how the strong fields decay into particles.
For this purpose, we first briefly review the Schwinger mechanism in the coexistence of electric and magnetic fields. We will point out that the vacuum immediately decays in the case of massless fermion and nonzero $\bm{E}$ and $\bm{B}$.
For simplicity, we consider the case that the electric and magnetic fields are covariantly constant~\cite{Suganuma}, i.e.,
$[D_\mu,\bm{E}]=[D_\mu,\bm{B}]=\bm{0}$, where $D_\mu=\partial_\mu-ig A_\mu$ is the covariant derivative with the gauge field $A_\mu$. The electric and magnetic fields are defined as
$\bm{E}^i=F^{i0}$ and $\bm{B}^i=-\epsilon^{ijk}F_{jk}/2$ with $F_{\mu\nu}=i[D_{\mu},D_{\nu}]/g$.
This is a generalization of constant fields in QED, 
$\partial_{\mu} \bm{E}=\partial_{\mu} \bm{B}=\bm{0}$, 
to the non-Abelian fields.
For the covariantly constant fields, all the components of 
$\bm{E}$ and $\bm{B}$ can be diagonalized to be constant matrices 
in color space by a gauge transformation.
Without loss of generality, 
one can also set $\bm{E}=(0,0,E)$ and $\bm{B}=(0,0,B)$ 
by choosing an appropriate Lorentz frame and the coordinate axis.
\section{Schwinger mechanism}
The vacuum decay in an electric field was discussed by
~\cite{Heisenberg,Schwinger}.
Consider the vacuum persistency probability, which is defined by
\begin{equation}
|\langle\Omega_\mathrm{out}|\Omega_\mathrm{in}\rangle|^2=\exp(-VTw) ,
\end{equation}
where $V$ and $T$ are infinite space volume and time length.
$|\Omega_\mathrm{in}\rangle$  and $|\Omega_\mathrm{out}\rangle$ are
the in-vacuum and the out-vacuum defined 
at $t=-T/2$ and $t=T/2$, respectively.
If the vacuum is unstable, $w$ has a nonzero value, 
while, if the vacuum is stable, $w$ vanishes.
Therefore, $w$ denotes magnitude of the vacuum decay 
per unit volume and time.
When $w$ is small, 
$|\langle\Omega_\mathrm{out}|\Omega_\mathrm{in}\rangle|^2\approx 1-VTw$, 
so that $w$ is regarded as the pair production probability 
per unit volume and time. 

For QCD, the analytic formula of $w$ for the quark-pair creation 
in the covariantly constant is given by~\cite{Suganuma}
\begin{equation}
  w=\sum_{n=1}^\infty \mathrm{tr} \frac{g^2 EB}{4\pi^2}\frac{1}{n}\coth(\pi n BE^{-1}) e^{-n\pi m^2/\sqrt{g^2E^2}} ,
\label{eq:wqcd}
\end{equation}
where $m$ denotes the quark-mass matrix and 
the trace is taken over the indices of color and flavor.
This is a non-Abelian extension of the following formula for QED~\cite{Tanji}:
\begin{equation}
  w=\sum_{n=1}^\infty \frac{e^{2} EB}{4\pi^2}\frac{1}{n}\coth(\pi n BE^{-1}) e^{-n\pi m^2/\sqrt{e^2E^2}} ,
\label{eq:w}
\end{equation}
with the QED coupling constant $e(>0)$.
Note that the fermion pair creation formalism 
in the covariantly constant fields in QCD is similar to that in QED, 
so that we hereafter give the formula for QED, where 
we set $E\ge 0$ and $B\ge 0$ 
by a suitable axis choice and the parity transformation. 

In the absence of the magnetic field, 
this formula reduces to the well-known result,
\begin{equation}
  w=\sum_{n=1}^\infty \frac{e^2 E^2}{4\pi^{3} }\frac{1}{n^2}
e^{-n\pi m^2/(eE)} .
\end{equation} 
If the masses are zero, $w$ has a  {\it finite} value of  $w= e^2 E^2/(24\pi)$.
The situation changes if the magnetic field exists. 
From Eq.~(\ref{eq:w}),  $w$ diverges in the presence of the magnetic field.
To see this, summing over all modes in Eq.~(\ref{eq:w}), we obtain for small $m$ as
\begin{equation}
  w\simeq  \frac{e^2EB}{4\pi^2}\ln\frac{eE}{\pi m^2}.
  \label{eq:wLLL}
\end{equation} 
As $m\to0$,  $w$ logarithmically diverges as
\begin{equation}
w \propto -\ln m\to \infty .
\end{equation}

Next, let us consider the origin of the divergence of $w$ in terms of 
effective dimensional reduction in a strong magnetic field. 
When a magnetic field exists, the spectrum of the transverse direction 
is discretized by Landau quantization. 
Actually, the energy spectrum for $E$=0 is given by
\begin{equation}
\varepsilon=\pm\sqrt{p_z^2+2eB(n +1/2\mp s_z)+m^2},
\label{eq:energy}
\end{equation}
where $n=0,1,\cdots$ correspond to 
the Landau levels, and $s_z=\pm1/2$ is the spin.
The system effectively becomes $1+1$ dimensional system 
with infinite tower of massive state: $m^2_{n, \mathrm{eff}}\equiv 2eBn+m^2$.
For the lowest Landau level (LLL), $n=0$ and $s=+1/2$, the energy is $\varepsilon=\pm\sqrt{p_z^2+m^2}$.
This is the spectrum in $1+1$ dimensions.
This LLL causes the divergence of $w$ as will be shown below.
The divergence of $w$ does not mean the divergence of 
the infinite pair production per unit space-time.
The divergence of $w$ rather implies that the vacuum always decays and produces pairs of fermion. 
The question is where the vacuum goes.
In the coexistence of $B$ and $E$, 
one can obtain the probability of the $n$ pairs of fermion with LLL as 
\begin{equation}
\begin{split}
&|\langle n \;\mathrm{ pairs}|\Omega_\mathrm{in}\rangle|^2 \\ 
&\quad= \exp\left[-\frac{V eB}{4\pi^2}\left(eET-\int dp_z n_{p_z}\right)\ln\frac{eE}{\pi m^2} \right].
\end{split}
\label{eq:numberOfPairs}
\end{equation}
The vacuum persistency probability corresponds to all $n_{p_z}$'s being zero in Eq.~(\ref{eq:numberOfPairs}), and
$w$ is equal to Eq.~(\ref{eq:wLLL}), so that $w$ diverges at $m=0$.
At $m=0$,
this probability is finite only if the following equation is satisfied:
\begin{equation}
eET-\int dp_z n_{p_z}=0 .
\label{eq:finiteProbabilityCondition}
\end{equation}
Therefore, the number of the particle with the LLL is restricted by Eq.~(\ref{eq:finiteProbabilityCondition}), and
linearly increases with time.
The higher Landau levels give heavy effective masses of order $eB$, so that all the contributions to the pair productions from
such modes are suppressed.
The total number of the particle pairs can be calculated:
\begin{equation}
N =VT\frac{e^2E^2}{4\pi^3}\frac{\pi B}{E}\coth\frac{\pi B}{E} .
\label{eq:numberOfPairsEB}
\end{equation}
At $B=0$, $N=VTe^2E^2/(4\pi^3)$.
The contribution of LLL is obtained as
\begin{equation}
N =VT\frac{e^2 E^2}{4\pi^3}\frac{\pi B}{E},
\end{equation}
which is equal to taking $\coth({\pi B}/{E})\to1$ in Eq.~(\ref{eq:numberOfPairsEB}).
\begin{figure}[tb]
   \centering
   \includegraphics*[width=0.9\linewidth]{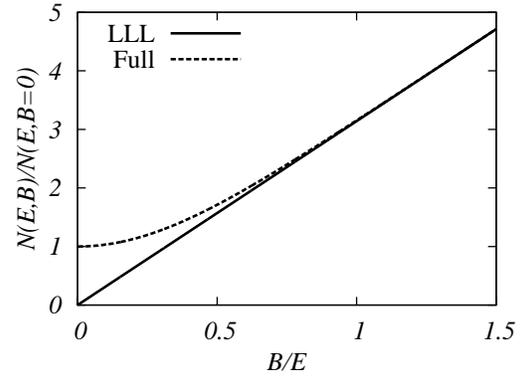}
   \caption{Ratio of the particle number to that at $B=0$. The solid line denotes the contribution from the LLL, and the dotted line denotes the contribution from all modes.}
   \label{fig:totalnumber}
\end{figure}
In Fig.~\ref{fig:totalnumber}, the total number of the particle for the full contribution and LLL contribution are shown. The LLL dominates for $B>E$, so that the effective model for the LLL works well for $B>E$. 

\section{Theory of strong magnetic field}
In this section, 
we study 
particle productions coming from the LLL for QED taken into account the back reaction.
For this purpose,  we consider LLL projected theory, that is, the wave function of the fermion is projected to 
the LLL state. The wave function of the LLL is
\begin{equation}
\begin{split}
\phi_l(x,y)&=\sqrt{\frac{eB}{2\pi l!}}\left(\frac{eB}{2}\right)^{\frac{l}{2}}(x+iy)^l\\
&\qquad\times\exp\left(-\frac{eB}{4}(x^2+y^2)\right) ,
\end{split}
\end{equation}
where $l$ denotes the angular momentum in $z$ direction for the LLL, 
and the energy is degenerate for $l$.
One can decompose the fermion field into the longitudinal mode and the transverse mode in a suitable representation as
\begin{equation}
\psi(x)= 
\begin{pmatrix}
\sum_l \phi_l(x,y)\varphi_l(t,z)\\
0 
\end{pmatrix} ,
\end{equation}
where $\varphi_l(t,z)$ is the two component Dirac field in 1+1 dimensions.
Then the fermion action of QED in $3+1$ dimensions reduces to that of non-Abelian gauge theory in $1+1$ dimensions:
\begin{equation}
\begin{split}
S&=\int d^4x \bar{\psi}(x)i\gamma^\mu D_\mu\psi(x)\\
&\simeq
\sum_{l,l'}\int dtdz \bar{\varphi}_{l'}(t,z)i\tilde{\gamma}^\mu \tilde{D}^{l'l}_\mu\varphi_{l}(t,z) ,
\end{split}
\label{eq:action}
\end{equation}
where $\tilde{\gamma}^t$ and $\tilde{\gamma}^z$ are the gamma matrices in $1+1$ dimensions and $\tilde{\gamma}^x=\tilde{\gamma}^y=0$.
The covariant derivative is defined by $\tilde{D}^{l'l}_\mu=\delta^{l'l}\partial_\mu-ie \tilde{A}^{l'l}_\mu$ with
\begin{equation}
\tilde{A}^{l'l}_\mu(t,z)=\int dxdy\phi_{l'}^*(x,y)\phi_l(x,y) A_\mu(x,y,z,t).
\label{eq:gaugeField}
\end{equation}
$\tilde{A}^{l'l}_\mu(t,z)$ corresponds to the gauge field in $U(\infty)$ gauge theory,
since $\tilde{A}^{l'l}_\mu(t,z)$ is an Hermite matrix, $\tilde{A}^{*l'l}_\mu(t,z)=\tilde{A}^{ll'}_\mu(t,z)$,
 and the indices $l$ and $l'$ run from $0$ to $\infty$.
To simplify the situation, we assume that the $A_t$ and $A_z$ do not depend on the transverse directions, $x$ and $y$.
Then the $l$ dependence can be factorized out: $\tilde{A}^{l'l}_\mu(t,z)=\delta_{ll'}\tilde{A}_\mu(t,z)$ and $\varphi_l(t,z)=\varphi(t,z)$.
The action in Eq.~(\ref{eq:action}) becomes
\begin{equation}
S\simeq\frac{eBV_\perp}{2\pi}\int dtdz \bar{\varphi}(t,z)i\tilde{\gamma}^\mu \tilde{D}_\mu\varphi(t,z) ,
\label{eq:action2}
\end{equation}
where $V_\perp$ is the volume of the transverse directions.
This action is nothing but that in 1+1 QED, i.e.,
the Schwinger model, except for the overall factor $eBV_{\perp}/(2\pi)$.
The exact solution of the effective action for the fermion is known as
\begin{figure}[tb]
   \centering
   \includegraphics*[width=0.9\linewidth]{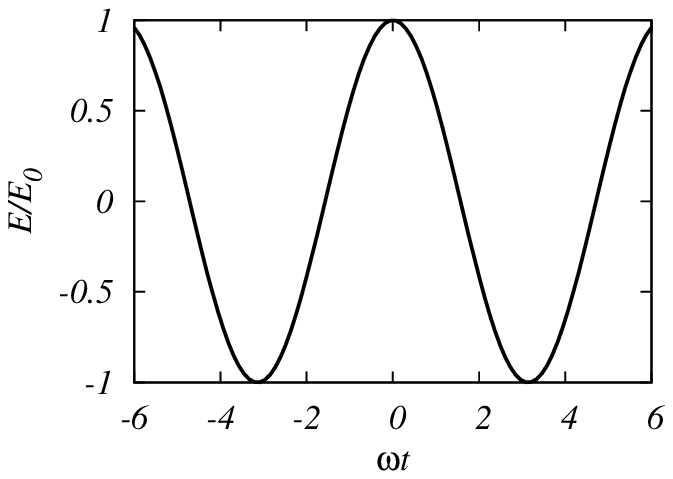}
   \centering
   \includegraphics*[width=0.9\linewidth]{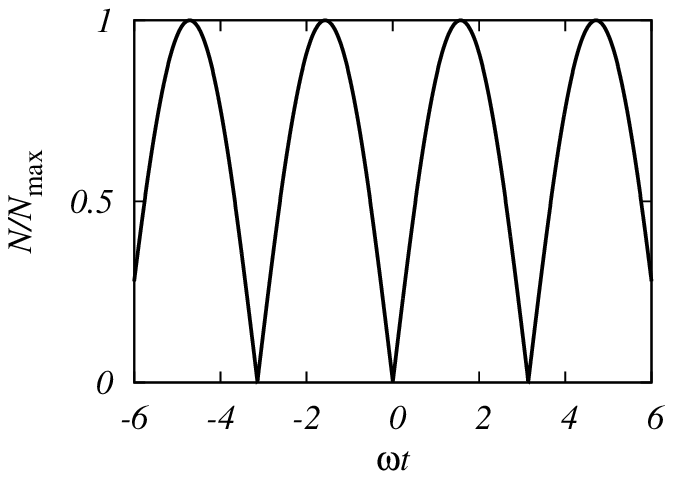}
   \caption{The electric field (upper) and the number of pairs (lower) at $z=0$ 
   plotted against time. Both values are normalized by that at maximum values.}
   \label{fig:dynamics}
\end{figure}
\begin{equation}
\begin{split}
\Gamma(A_\mu)&= -\frac{m_\gamma^2V_\perp}{2}\int dtdz \tilde{A}_{\mu}(t,z) \\
&\qquad\times\left(g_\parallel^{\mu\nu}-\frac{\partial_\parallel^\mu\partial_\parallel^\nu}{\partial_{\parallel}^2}\right)\tilde{A}_\nu(t,z),
\end{split}
\label{eq:effectiveAction}
\end{equation}
where $\parallel$ denotes for longitudinal directions, $t$ and $z$.
$m_\gamma$ denotes the effective photon mass,
$m_\gamma^2\equiv e^3B/(2\pi^2)$.
Equation~(\ref{eq:effectiveAction}) is manifestly gauge invariant. 
In the Lorenz gauge,  it reduces to the result by \cite{Schwinger3}.
The mass $m_\gamma$ is induced by the axial anomaly, of which effect is called ``dynamical Higgs effect.''
This mass generation is related to the fact $w\to\infty$ as $m\to0$.
Using this form, we can calculate the fermion and axial currents,
\begin{eqnarray}
j^\mu(x)\!\!\!&=&\!\!\!\frac{\delta\Gamma(A)}{\delta (eA_\mu)}=-\frac{e^{2}B}{2\pi^{2}}\tilde{A}^\mu(t,z), \\
j_5^\mu(x)\!\!\!&=&\!\!\!-\epsilon^{\mu\nu}j_\nu(x),
\end{eqnarray}
where we choose the Lorenz gauge, $\partial_\mu \tilde{A}^\mu=0$.
The divergence of the axial current leads the axial anomaly in $1+1$ dimensions except for the overall factor $eB/(2\pi)$:
\begin{equation}
\partial_\mu j_5^\mu(x)=\frac{e^2B}{2\pi^2}\epsilon^{\mu\nu}\partial_{\mu}\tilde{A}_{\nu}(t,z)
=\frac{e^2}{2\pi^2}BE.
\label{eq:anomaly}
\end{equation}
This relation is nothing but axial anomaly in $3+1$ dimensions.
Since the effective action in Eq.~(\ref{eq:effectiveAction}) 
has a quadratic form in $\tilde{A}_{\mu}$, 
the equation of motion for the photon can be solved. 
For example, the electric field of $z$ direction is
\begin{equation}
E=E_0\cos(\omega t-k_zz ) ,
\end{equation}
where $\omega=\sqrt{k_z^2+m_\gamma^2}$. The currents satisfy
$ej^\mu=-\epsilon^{\mu\nu}\partial_\nu E$ and
$ej_5^\mu=\partial^\mu E$.
We show the electric field and the number density for the spatially homogeneous case, $k_z=0$,
as a function of time in Fig.~\ref{fig:dynamics}. 
The electric field oscillates with a frequency $\omega$.
In this case, 
$j^{t}=0$, 
but 
$j^{t}_5\neq0$.
The number density of pairs is equal to $|j_5^t|/2$.
These results agree with the previous works~\cite{Tanji,Iwazaki}.

The generalization to non-Abelian theories is straight forward if the magnetic field is enough strong. The fermion determinant becomes Wess-Zumino-Witten action.

\section{Summary and Outlook}
In this work, we have discussed the vacuum decay in strong electric and magnetic fields.
When the fermion is massless, 
the vacuum persistency probability per unit time and volume becomes zero, 
and hence $w$ diverges. The origin of the divergence is from discretized spectra of transverse directions and the lowest Landau level. The LLL level dominates for $B>E$.
With the LLL projection, 
we have analytically calculated the effective action in this situation, and reproduced the previous numerical results.
Since the effective theory of the LLL is solvable, 
there is no chaotic behavior nor thermalization. The thermalization does not happen
in the LLL-dominant process.

In the case of QCD, the gluon is more interesting because of its self-interaction.
The Landau quantization causes instability because the helicity of gluon is one.
Inserting $s_z=1$ in Eq.~(\ref{eq:energy}), the energy becomes imaginary when $p_z<eB$,
of which instability is known as Nielsen-Olesen instability.
The situation is similar to quenching
phenomenon in general phase transition, where a temperature suddenly changes.
In such a situation, a phase separation occurs. 
The same phenomena would happen in relativistic heavy-ion collisions,
because the electric and magnetic fields are suddenly induced by the collision, and 
the perturbative vacuum of gluon is unstable.
Although dynamics of unstable gluon vacuum is very interesting,
we leave this topic in the future work.

\end{document}